\documentclass[fleqn]{aa}

\usepackage{amsmath}
\usepackage{graphicx}
\usepackage{txfonts}
\usepackage{enumitem}
\usepackage[square]{natbib}

\begin{document}
\mathindent=0pt
%Opening
\title{Full Stokes polarimetric observations with a single-dish radio-telescope}

   \author{E. Cenacchi
          \inst{1}\thanks{E.C. is a member of the International Max Planck Research School for Radio and Infrared Astronomy. E-mail: cenacchi@mpifr.de}
	   A. Kraus
	  \inst{1}
           A. Orfei
          \inst{2}
	  \and
           K.-H. Mack
          \inst{2}}

   \institute{Max-Planck-Institut f\"ur Radioastronomie, Auf dem H\"ugel 69, D-53121 Bonn, Germany
         \and
           Istituto di Radioastronomia, INAF, Via P. Gobetti 101, I-40129  Bologna, Italy}

   \date{Received 22 October 2008 / Accepted 22 January 2009}

\abstract
%Context
{The study of the linear and circular polarization in AGN allows one to gain detailed information about the properties of the magnetic fields in these objects. However, especially the observation of circular polarization (CP) with single-dish radio-telescopes is usually difficult because of the weak signals to be expected. Normally CP is derived as the (small) difference of two large numbers (LHC and RHC); hence an accurate calibration is absolutely necessary.}
%Aims
{Our aim is to improve the calibration accuracy to include the Stokes parameter V in the common single-dish polarimetric measurements, allowing a full Stokes study of the source under examination.}
%Methods
{A detailed study, up to the 2nd order, of the M\"uller matrix elements in terms of cross-talk components allows us to reach the accuracy necessary to study circular polarization.}
%Results
{The new calibration method has been applied to data taken at the 100-m Effelsberg radio-telescope during regular test observations of extragalactic sources at 2.8, 3.6, 6 and 11 cm. The D-terms in phase and amplitude appear very stable with time and the few known values of circular polarization have been confirmed.}
%Conclusions
{It is shown that, whenever a classical receiver and a multiplying polarimeter are available, the proposed calibration scheme allows one to include Stokes V in standard single-dish polarimetric observations as difference of two native circular outputs.} 

\authorrunning{E. Cenacchi et al.}
\maketitle

%Chapter 1
\section{Introduction}
The study of circular polarization (CP) and accurate multi-frequency observations of linear polarization (LP) provide probes of AGN magnetic field properties. CP, LP and spectral information can be used to constrain the low energy end of the relativistic particle distribution \citep{Beckert}, derive magnetic field order, strength and geometry \citep{Gabuzda} and make inferences about the particle content of the AGN jets \citep{Wardle}. So far several studies aimed at studying CP from AGN have been carried out with interferometers: ATCA at 5 GHz \citep{Rayner}, VLBA at 5.0 GHz and 15.0 GHz \citep{HomanWardle} and the VLA at 8.4 GHz \citep{Bower}. On the other hand for instance the Michigan 26-metre radio-telescope at 4.8 GHz and 8.0 GHz \citep{Allers} has been used to perform single-dish CP observations.

Stokes $V$, which measures CP through the ratio $V/I$, has often been discarded from the polarimetric observation strategy and calibration, due to its typically low values, leading to a strong preference for radio-astronomical receivers to provide circularly polarized outputs, as this configuration allows efficient measurement of Stokes $I, Q$ and $U$.

When a single-dish telescope delivers two opposite circular outputs, it is relatively easy to retrieve information about Stokes $I$ (sum of the outputs) as well as $Q$ and $U$ (cross-correlation of the outputs), but the estimate of Stokes $V$ (difference of the outputs) can be severely contaminated by the receiver internal noise and gain fluctuations and by small intrinsic asymmetries in the receiver channels structure (e.g. differences between cables lengths, different internal gains, unwanted internal reflections). Some of these effects cause a small percentage of contamination between the signals belonging to the left and the right channel (D-terms), in amplitude and phase.

To fill the gap that Stokes $V$ leaves in the single-dish polarimetric observations, the most common receiving architecture has been examined. A full Stokes calibration scheme has been derived by identifying at each step the possible contamination that may be introduced by the devices used in the receiving chain, along with the dependance of the 4$\times$4 M\"uller matrix elements on the D-terms, up to 2nd order. The result has been applied to several observations carried out at the Effelsberg 100-m telescope.

The application of an accurate 4$\times$4 M\"uller matrix to the observed raw data offers the possibility of performing multi-frequency full Stokes observations of the sources, giving a complete description of their polarization spectra.

%Chapter 2
\section{Basics of polarimetry}
% Ch. 2.1
\subsection{Polarized waves}

By focusing on the electric field only (the magnetic field can always be derived from $ \overrightarrow{E}=-\overrightarrow{c} \times \overrightarrow{B} $) a wave that is travelling along the $z$-axis with complex amplitude $\widetilde E_0$ can always be decomposed into two components, one along the $x$-axis and one along the $y$-axis, as follows:
\begin{align} \label{classpol}
\overrightarrow{E} & = \hat{\imath}E_{x}e^{i(kz-\omega t + \delta_{x})}+\hat{\jmath}E_{y}e^{i(kz-\omega t + \delta_{y})} \nonumber \\
& = \left[\hat{\imath}E_{x}e^{i \delta_{x}}+\hat{\jmath}E_{y}e^{i \delta_{y}}\right] e^{i(kz- \omega t)} \\
& = \widetilde E_0e^{i(kz-\omega t)} \nonumber
\end{align}
where $\hat{\imath}$ and $\hat{\jmath}$ are unit vectors in the $x$ and $y$ direction.
\\

Polarization is a phenomenon which can occur in transversal waves and is described by the ``trace'' left by a wave passing through a plane perpendicular to the propagation direction. A well defined polarization state (linear, circular, elliptical, also called ``mode'') can be recognized in all those cases where a constant relation between $E_{x},E_{y}$ and $\delta _{x}, \delta _{y}$ exists and reflects the existence of a defined ordered motion of the electric vector.

Any polarized wave can be considered as composed of two linearly polarized waves, as shown in Eq. \eqref{classpol}, but any pair of orthogonally polarized waves (e.g. left circular/right circular, left elliptical/right elliptical with axis at 90$^{\circ}$, etc.) can be used as a basis for description of a polarization state. The most convenient choice is typically the polarizations to which the telescope outputs respond.

%Ch. 2.2
\subsection{Jones vector and  M\"{u}ller matrix in the LR circular representation}
The polarization properties of the radiation can be described using different vectors. The Jones vector describes the information related to amplitude and phase of the observed signal components, thus it is the preferred one to describe mathematically the receiver internal devices. The Stokes vector reflects the relation between the signal components and defines the overall polarization of the signal, thus it is the preferred one to describe the source physical characteristics. 

The Jones and Stokes vectors carry the same information and are linked by a set of transformation equations. As a rule of thumb, the basis for the transformation should be chosen according to the polarization of the outputs delivered by the telescope under examination. The following equations are related to the circular LR frame, which we adopted in this paper following \citet{Kraus} and \citet{Heiles}. 

The Jones vector, $\overrightarrow{J}$, and the Stokes vector, $\overrightarrow{S}$, are defined as follows

\begin{equation} \label{jones}
\overrightarrow{J}=\left[ \begin{array}{c} 
\widetilde E_L \\ \widetilde E_R \end{array} \right] =\left[\begin{array}{c} 
E_L e^{-i\delta_{L}} \\ E_R e^{-i\delta_{R}} \end{array}\right]
\end{equation}
\begin{equation} \label{stokes}
\overrightarrow{S}=\left[ \begin{array}{c} 
I \\ Q \\ U \\ V \end{array} \right]
\end{equation}

\begin{subequations} \label{stokesjones}
\begin{align} 
& I = \left\langle E_L^2\right\rangle +\left\langle E_R^2\right\rangle=\widetilde E_L^*\widetilde E_L+\widetilde E_R^*\widetilde E_R  \\
& Q = 2\left\langle E_L E_R\cos\Delta\right\rangle=\widetilde E_L \widetilde E_R^* + \widetilde E_R \widetilde E_L^* \\ 
& U = 2\left\langle E_L E_R\sin\Delta\right\rangle=i\left(\widetilde E_R \widetilde E_L^*-\widetilde E_L \widetilde E_R^*\right) \\
& V = \left\langle E_L^2\right\rangle -\left\langle E_R^2\right\rangle=\widetilde E_L^*\widetilde E_L-\widetilde E_R^*\widetilde E_R 
\end{align}
\end{subequations}
where the asterisk stands for complex conjugate and $\Delta=\delta_R-\delta_L$.
\\

The M\"{u}ller matrix relates the polarimetric properties of the source, i.e. the Stokes parameters, of the source with the measurements from the polarimeter at the telescope. Hence it describes the instrument polarization characteristics. An ideal instrument that measures Stokes parameters perfectly, would be characterized by a unitary M\"uller matrix $\overrightarrow{S}_{m}=\mathbf{M}\overrightarrow{S}_{s}$
\begin{equation} \label{mueller1}
\mathbf{M}=\left(\begin{array}{cccc} 
m_{11} & m_{12} & m_{13} & m_{14} \\ 
m_{21} & m_{22} & m_{23} & m_{24} \\ 
m_{31} & m_{32} & m_{33} & m_{34} \\
m_{41} & m_{42} & m_{43} & m_{44} \end{array}\right) \\
\end{equation}

\begin{subequations} \label{mueller2}
\begin{align} 
& I_{m}=m_{11}I_{s}+m_{12}Q_{s}+m_{13}U_{s}+m_{14}V_{s} \\
& Q_{m}=m_{21}I_{s}+m_{22}Q_{s}+m_{23}U_{s}+m_{24}V_{s} \\
& U_{m}=m_{31}I_{s}+m_{32}Q_{s}+m_{33}U_{s}+m_{34}V_{s} \\
& V_{m}=m_{41}I_{s}+m_{42}Q_{s}+m_{43}U_{s}+m_{44}V_{s} 
\end{align}
\end{subequations}
where $I_{m}$, $Q_{m}$, $U_{m}$ and $V_{m}$ are the measured Stokes parameters and $I_{s}$, $Q_{s}$, $U_{s}$ and $V_{s}$ are the true Stokes parameters describing the source polarization.
\\

When handling linear polarization, the polarization angle is measured with respect to some instrumental zero point, usually related to a reference direction on the celestial sphere. During the source tracking, the reference system of an alt$-$azimuth mounted telescope rotates with respect to the source, that is, the parallactic angle of the sources (and consequently its polarization angle) rotates during the observation. Following \citet{Turlo} the time$-$dependent rotation component $\mathbf{B}$ of the M\"{u}ller matrix can be extracted and applied separately from the instrumental component of the M\"{u}ller matrix, $\mathbf{T}$ is assumed to be constant with time. The overall M\"{u}ller matrix can be then expressed as $\mathbf M = \mathbf{TB}$, where $\mathbf{B}$ is the rotation matrix of a linear rotating system (rotation of $\vartheta$)

\begin{equation} \label{rotation}
\mathbf{B}=\left( \begin{array}{cccc} 
1 & 0 & 0 & 0 \\ 
0 & \cos 2\vartheta & -\sin 2\vartheta & 0 \\ 
0 & \sin 2\vartheta & \cos 2\vartheta & 0 \\ 
0 & 0 & 0 & 1 \end{array}\right), \quad
\mathbf{B^{-1}}=\left( \begin{array}{cccc} 
1 & 0 & 0 & 0 \\ 
0 & \cos 2\vartheta & \sin 2\vartheta & 0 \\ 
0 & -\sin 2\vartheta & \cos 2\vartheta & 0 \\ 
0 & 0 & 0 & 1 \end{array}\right).
\end{equation}

%Chapter 3
\section{Application to a single-dish observation}
\begin{figure*}
\caption{Receiver sketch. The block diagram shows the main parts of a typical radio-astronomical receiver: the scalar feed, sensitive to all the polarization states; the directional coupler, that couples the linearly polarized noise diode signal into the source radiation; the polarizer, that transforms the signal into the circularly polarized representations; the OMT, that splits the orthogonally polarized signals (polarizer and OMT are sometime realized as a single device); the low noise amplifiers (LNA); the down-conversion parts, here represented in a single block (labeled ``conversions''); the total power detectors (TP); the two multipliers, here shown as a single device, that take as input the LHC and RHC channels to give out Stokes Q, and the LHC and the 90$^\circ$ phase shifted RHC to give out U. The picture shows also equations giving the electric field strength, or voltage in the parts where propagation is as current on a wire, how those depend on the gain and the D-terms factors and how they are combined during each propagation step, see Ch. 4.1 and Ch. 4.2 for a complete derivation.} 
\label{p:receiver}
\centering
\includegraphics[width=520 pt]{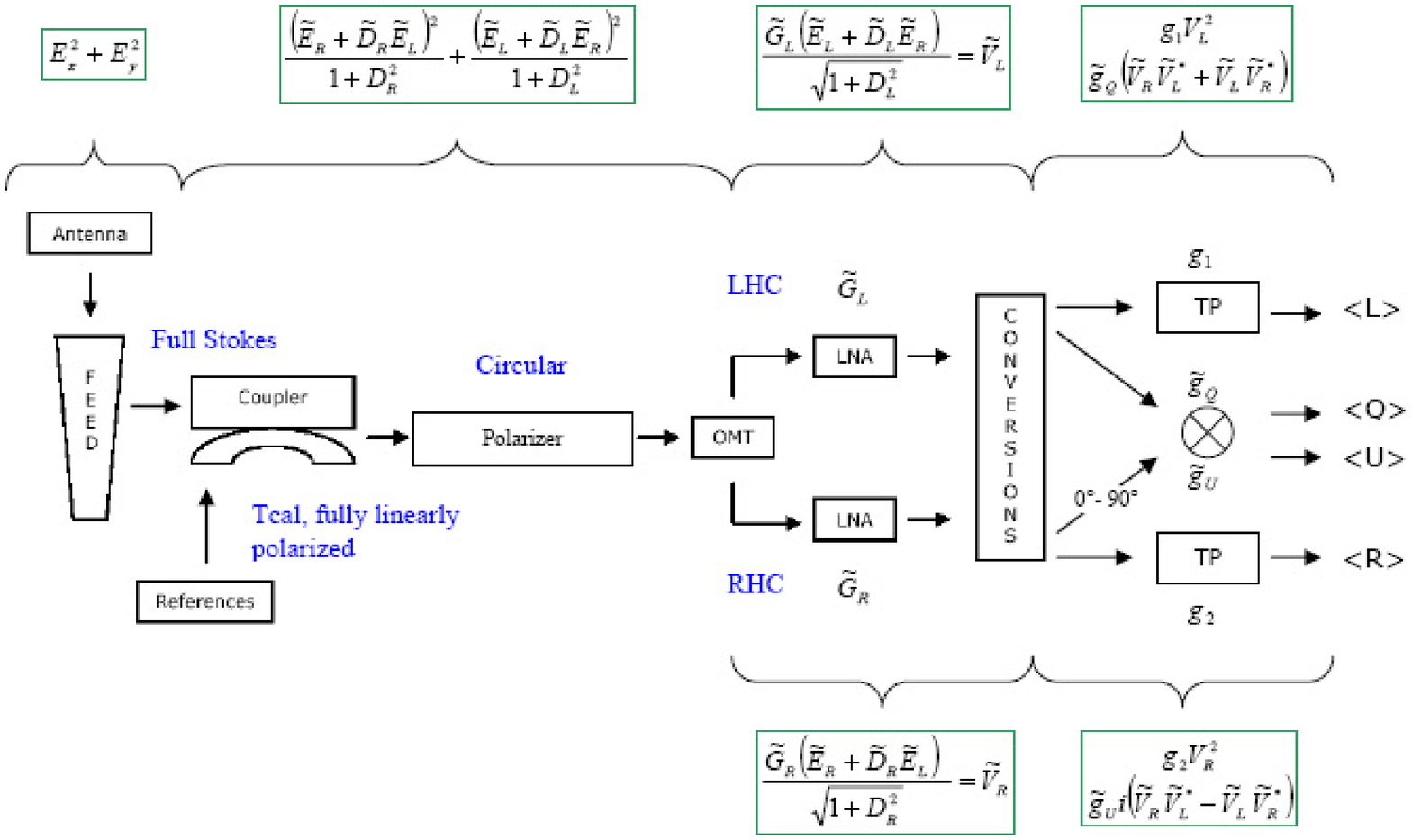}
\end{figure*}

\citet{McKinnon} and \citet{Hamaker1} published major works on how to manage polarimetry with a radio interferometer and \citet{Johnston} adapted their study to single-dish antennas equipped with linear dipoles. Nowadays telescopes are commonly equipped with scalar feeds and hybrid or wave-guide elements that supply circularly polarized outputs. In the following we describe, step by step, how this instrumentation can affect, in terms of instrumental polarization, the measurement of the incoming radiation. A typical receiving chain is sketched in Fig. \ref{p:receiver}. In Table \ref{t:notation} we summarize our notation.

\begin{table}[htbp]
\caption{Here the notation adopted to refer to different quantities is summarized. All complex quantities are expressed in the form $\widetilde{A}=Ae^{ip}$, where $A$ is the amplitude and $p$ is the phase. $P$ is the phase difference between the two channels.}
\label{t:notation}
\centering
\begin{tabular}{lccc}
\hline
 & $A$ & $p$ & $P$ \\
\hline
Incoming radiation $\widetilde E $ & $E_R, E_L $ & $\delta_R,\delta_L $ & $\Delta=\delta_R-\delta_L$ \\
Cross term $\widetilde D $ & $D_R ,D_L $ & $\varphi_R,\varphi_L$ & $\Phi=\varphi_L-\varphi_R$ \\
Receiver gain $\widetilde{G}$ & $G_R,G_L$ & $\psi_R,\psi_L$ & $\Psi=\psi_L-\psi_R$ \\
Backend gain: Total Power $g$ & $g_1,g_2$ & $-$ & $-$ \\
Backend gain: Polarimeter $\widetilde g$ & $g_Q,g_U$ & $\gamma_Q,\gamma_U$ & $-$ \\
Outgoing signal $\widetilde V$ & $V_R,V_L$ & $-$ & $-$ \\
\hline
\end{tabular}
\end{table}

%Ch.3.1
\subsection{The antenna}
The targets of single-dish observations are typically point-like with respect to the antenna beam and the observables are the integrated Stokes parameters and the full Stokes spectra of the source. The local CP and LP can be observed only resolving the source with interferometric techniques, and are often averaged out when the source is not spatially resolved (beam depolarization).

An ideal rotationally symmetric telescope does not change the state of polarization of the incident radiation (though circular polarization changes its state every time it is reflected, due to changing direction of propagation) see \citet{McGuire}. 

The first deviation from the ideal case is due to the blockage by the subreflector and its supporting structure. The radiation scattering causes a cross-polarization effect, both in amplitude and phase, which can be widely reduced by the usage of an appropriate structure design. Coupling the antenna with corrugated horns or dual mode horns with circularly symmetric patterns offers cross-polarization efficiencies closer to 1.

The parabolic shape is usually optimized at a well defined elevation, typically around 45$^{\circ}$. When the antenna observes far from this value, its shape deviates from the original parabola (and the symmetry could be broken), unless it is equipped with some correction mechanism that compensates the gravitational deformations (active surfaces or homology). As an example of the deformation involved, Table \ref{t:Medicina} lists the deviation from the best-fit parabola for a typical dish without compensating techniques.

\begin{table}[htbp]
\caption{Deviation from the best fit parabolic surface, typical values}
\label{t:Medicina}
\centering
\begin{tabular}{ccc}
\hline
Elevation & Deviation &  Deviation \\
$[^{\circ}]$ & [rms mm] & [\% @ 6 cm]\\
\hline
30 to 60 & 0.5 & 0.8 \\
20 to 75 & 0.8 & 1.3 \\
\hline
\end{tabular}
\end{table}
%Ch. 3.2
\subsection{The feed}
The feed is used to obtain a Gaussian beam pattern and converts the electromagnetic field from propagating in free space to propagating in waveguide. The most common type, coupled to Cassegrain and Gregorian antennas, is a conical corrugated horn (often called ``scalar feed''), as it provides a Gaussian beam profile and is characterized by negligible cross-polarization. Unlike linear dipoles or helical antennas, the scalar feed usually does not have a single defined polarization mode, as it can support linear, circular and dual polarization, depending on its design and on the electronic devices that are coupled to it. A typical scalar feed provides a single output: some additional devices are needed to split it in the polarized components.

A lack of symmetry in the feed conical shape could lead to some degree of polarization aberrations. As an example of deviation from the ideal feed shape, the typical mechanical accuracy is 0.02 mm to 0.03 mm. This would lead, at 6 cm (nearly 600 mm of diameter, in secondary focus) to a possible non-circularity of 0.003\% to 0.005\%.

For off-axis feeds (multibeam or focal plane receivers that permanently hosts several different feeds) the rotational symmetry is broken. As a result, a certain amount of instrumental polarization enters and increases the polarized signal $I_{p}$ and the antenna patterns for the two orthogonal polarization channels of the receiver are different (beam squint). The result is a constant pointing offset between the two opposite beams which depends on the distance of the feed from the optical axis \citep{Fiebig}.

%Ch. 3.3
\subsection{The directional coupler for calibration}
A radio-astronomical receiver suffers from a certain amount of gain instability, due to electronic noise added by each component, thus two measurements of an identical power source could yield slightly different voltage values. To stabilize this fluctuation an internal noise of known temperature (Tcal) is added to the astronomical signal using a directional coupler. This calibration signal is linearly polarized and is injected into the receiver with a known polarization angle. The Tcal is also used to convert the measured voltage into kelvin.

The value of the Tcal is usually known with an accuracy of 3\% to 5\% over a wide bandwidth (e.g. 500 MHz for the 6 cm receiver at the Effelsberg telescope). 

A typical value of the stability is of 0.01 dB$/$K, this number must be interpreted as a Tcal variation of 0.2\% per each 1 K variation in the receiver room temperature (which is usually kept constant). Assuming a Tcal nominal value of 2 K and a thermal excursion inside the room of $\pm$ 5 K a value of (2.00$\pm$0.02)K is obtained.

%Ch. 3.4
\subsection{The polarizer and the ortho-mode transducer (OMT)}
The polarizer and the OMT split the signal in two orthogonal circular polarized components. This process is not perfect and some fraction of one component can affect the other component. This effect is known as ``cross-talk'' and is quantified by the so called ``D-terms'' (values $<$ 1\% are usually considered very good).

The ensemble of feed, polarizer and OMT is completely analogous to a pair of circularly polarized feeds, although the instrumental effects on the polarization measurements are different. 

At this point in the receiving chain the following can be expected:

\begin{itemize}[leftmargin=*]
\item Antenna: the non-perfect rotational symmetry can cause diattenuation and birefringent effects (that is, some instrumental polarization and some conversion among $Q, U, V$). For circular polarization, a change in sign must be expected at each reflection. These effects are usually cancelled by appropriate technical equipment (active surface, appropriate cabling to compensate the change in sign) and by a calibration that compensates possible residual components (e.g. antenna gain curve as function of elevation).\\

\item Feed: small mechanical imperfections cause additional diattenuation and birefringent effects, but at low frequencies this should be a negligible contribution, as, e.g. at 6 cm, the imperfections are typically of the order of 0.01\%.\\ 

\item Directional coupler: the phase of the transmitted signal is not constant over the bandwidth, this typically causes an additional cross-polarization.\\

\item Polarizer and OMT: a small part of the left circularly polarized signal could enter the right circularly polarized signal, and vice versa. This could heavily affect the $V$ measurement while the effect on $I, Q, U$ should be negligible.\\

\item LNA: a small fraction of the received radiation is reflected back to the previous components due to inevitable impedance mismatches, an effect called non-zero return loss. Once this spurious signal passes through the polarizer it faces the directional coupler and is reflected back again. The two reflections added to a passage through the polarizer transform the circular polarization into its orthogonal polarization. This leads to a spurious exchange of radiation between the left and right-hand channels.
\end{itemize}
%Ch. 3.5
\subsection{The down-conversion}
After the OMT, the two circularly polarized signals are amplified and downconverted. The ``conversion box'' can be considered linear (some deviation appears for very strong sources) but could introduce some polarization aberration: some small difference in the signal path length could alter the phase difference between the two components, and the amplitude could also be slightly modified. The signal, after the front-end according to \citet{Conway} and \citet{McKinnon}, is then (see Table \ref{t:notation} for the adopted notation)
\begin{subequations} \label{conway}
\begin{align} 
\widetilde V_R & = \frac{\widetilde G_R\left(\widetilde E_R+\widetilde D_R\widetilde E_L\right)}{\sqrt{1+D_{R}^2}} \\
\widetilde V_L & = \frac{\widetilde G_L\left(\widetilde E_L+\widetilde D_L\widetilde E_R\right)}{\sqrt{1+D_{L}^2}} 
\end{align}
\end{subequations}

\noindent where $\widetilde G_R=G_R e^{-i\psi_R}$ and $\widetilde G_L=G_L e^{-i\psi_L}$ are the gain of the right and left channels and $\widetilde D_R=D_R e^{-i\varphi_R}$ and $\widetilde D_L=D_L e^{-i\varphi_L}$ are the cross$-$talk factors.

The receiver architecture usually offers narrow bandwidths (some tenths of MHz) at low frequencies ($\leq$ 3 GHz) and wider bandwidths (some hundreds of MHz, few thousands in some new generation receivers) at higher frequency ($>$3 GHz). It can be shown that under these circumstances the bandwidth depolarization factor (see \citet{Gardner}), which affects LP and is defined as the ratio between the observed degree of linear polarization and the emitted one, is negligible for typical extragalactic sources (a typical Rotation Measure of 50 $rad/m^{2}$ gives a depolarization factor of nearly 1 both at 2.5 GHz over a 100 MHz band and at 5 GHz over a 500 MHz band).

%Ch. 4
\section{Correction of observations using the instrumental M\"{u}ller matrix, T}
After the down-conversion, the signals are sent to the devices that perform the measurements by implementing Eqs. \eqref{stokesjones}. Typical devices are square law detectors (to obtain the flux densities from the left and right channels) and multiplying polarimeters (which supply $Q$ and $U$, through a 90$^{\circ}$ shift of $V_L$ or $V_R$). In App. \ref{App1} the calibration procedure is described (the complete calculations are detailed in Cenacchi (2009, PhD thesis, in prep.).
%Ch.4.1
\subsection{Effects on the measurements}
\noindent The instrumental M\"{u}ller matrix elements contain information about the amount of contamination among the four Stokes parameters, due to spurious, unwanted, conversions arised inside the receiver. The conversion of $Q, U$ and $V$ into $I$ is typically negligible, the conversion of $I$ and $V$ into $Q$ and $U$ represents a small but still significant corruption, the conversion of $I, Q$ and $U$ into $V$ can completely overwhelm the result of the measurement, leading to the impossibility of disentangling the instrumental contribution from the intrinsic $V$ value. This clearly appears when dealing with completely unpolarized (thermal) sources. 

As an example, Table \ref{t:rawpol} lists the Stokes values measured in November 2007 during a test observation with the Effelsberg 100-m telescope at 6 cm of two planetary nebulae. Being thermal sources they are expected to be unpolarized. The sources were observed with five beam wide on-the-fly cross-scans centred on the source. The $Q$ and $U$ values came directly from the multiplying polarimeter. The $V$ value has been estimated as the difference between the left and right channels, after those were separately corrected for pointing (to compensate for the beam squint) and gain curve. The cross-scan technique intrinsically removes possible differences between the internal channel noise ($\approx$ 1 K for the Effelsberg telescope at 6 cm).

\begin{table*}[htbp]
\caption{Examples of raw measurements of all Stokes parameters for unpolarized sources at Effelsberg at 6 cm. Measurements were calibrated using a noise diode of known strength, but D-terms correction was not applied.}
\label{t:rawpol}
\centering
\begin{tabular}{lrrrrrrrrrrrr}
\hline
Source & $I_m$ & $\sigma_{\Delta}$ & $Q_m$ & $\sigma_{\Delta}$ & $U_m$ & $\sigma_{\Delta}$ & $V_m$ & $\sigma_{\Delta}$ & LP & $\sigma_{\Delta}$ & CP & $\sigma_{\Delta}$  \\
 & [K] & & [K] & & [K] & & [K] & & [\%] & & [\%] & \\
\hline
NGC\,6572 & 3.861 & 0.008 & -0.016 & 0.002 & 0.006 & 0.013 & -0.016 & 0.004 & 0.43 & 0.12 & -0.42 & 0.11 \\
NGC\,7027 & 17.100 & 0.300 & -0.076 & 0.011 & -0.014 & 0.015 & -0.087 & 0.022 & 0.45 & 0.07 & -0.51 & 0.13 \\
\hline
\end{tabular}
\end{table*}

The measured $Q, U$ and $V$ values, from unpolarized sources, are believed to be due to a contamination from Stokes $I$, that is, from a small net exchange in amplitude and phase between the left and right channels that, despite the multiplication and the correction for pointing and gain, still produces a spurious significant LP and CP detection.

App. \ref{App2} contains the definition of each element of the M\"uller matrix.

%Ch. 5.3
\subsection{How to apply the D-terms calibration}
\subsubsection{Derivation of the M\"uller matrix}
The proposed calibration scheme is completely based on observation of astronomical calibration sources, each observation schedule must include a fully unpolarized source ($Q=U=V=0$) and a highly linearly polarized source (for which one does not require $V=0 K$).
The left and right channels must be corrected separately for pointing and gain curve. The raw $V$ value can then be estimated as the difference of the measured kelvin. R and L opacity correction can be applied to the channels and the raw $I$ can be derived as the sum of the R and L  temperatures.
The raw $Q$ and $U$ come directly from the multiplying polarimeter.

By observing one or more unpolarized sources whose flux density is known, Eq. \eqref{mueller2} is simplified as follows
\begin{subequations} \label{muellersimpl}
\begin{align} 
& I_{m}=t_{11}I_{s} \\
& Q_{m}=t_{21}I_{s} \\
& U_{m}=t_{31}I_{s} \\
& V_{m}=t_{41}I_{s} 
\end{align}
\end{subequations}

where $t_{11}, t_{21}, t_{31}$ and $t_{41}$ are given in Eqs. \eqref{dlist}. These four matrix elements involve the parameters $D_L, D_R, \varphi_L, \varphi_R, \Delta_c$ and $\Psi$, all of which must be known for the calibration to be applied. 

The polarization angle of the internal noise signal, $\Delta_c$ is known (either from technical specifications or from previous measurements) and the remaining five quantities can be solved from six measurements on the calibrators (four Stokes parameters measurements on the unpolarized source and Stokes $Q$ and $U$ measurements on the linearly polarized calibrator).

Once the D-terms, in amplitudes and phases, are determined, all terms of $\mathbf{T}$ can be calculated, as they depend on only these parameters. 
By applying this procedure, it is not necessary to make assumptions about the $V$ value from the linearly polarized calibrator.

%Ch. 5.3.2
\subsubsection{Treatment of the errors}

Let the uncertainties measured on $I_m, Q_m, U_m$ and $V_m$ for the calibrator(s) be $\sigma_I, \sigma_Q, \sigma_U$ and $\sigma_V$, then the corresponding uncertainty on the matrix elements are
\begin{subequations} \label{Errors1}
\begin{align}
&\sigma_{11} = \left\lvert \frac{\sigma_I}{I_s}\right\rvert\\
&\sigma_{21} = \left\lvert \frac{\sigma_Q}{I_s}\right\rvert\\
&\sigma_{31} = \left\lvert \frac{\sigma_U}{I_s}\right\rvert\\
&\sigma_{41} = \left\lvert \frac{\sigma_V}{I_s}\right\rvert
\end{align}
\end{subequations}

Due to the non-linearity of the equations involved, it is not possible to propagate these errors through the estimate of the cross-terms, to the complete instrumental M\"uller matrix. It is anyway possible, by comparing the mathematical definitions of the matrix elements, to use the results \eqref{Errors1} to establish an upper limit for the remaining errors, e.g. it is clear from their definition that the error on $m_{11}$ and $m_{14}$ must be the same. By applying similar qualitative considerations, the complete error matrix can be estimated as

\begin{equation} \label{Errors2}
\mathbf{\sigma_T}=\left(\begin{array}{cccc} 
\sigma_{11} & \sigma_{11} & 3\sigma_{11} & \sigma_{11} \\ 
\sigma_{21} & \sigma_{21} & \sigma_{21} & \sigma_{21} \\ 
\sigma_{31} & \sigma_{21} & \sigma_{21} & \sigma_{31} \\ 
\sigma_{41} & \sigma_{11} & 3\sigma_{11} & \sigma_{41} \end{array}\right).
\end{equation}

The intrinsic Stokes vector can be then obtained by inverting $\mathbf{T}$, propagating step by step the errors $\sigma_{nm}$ during the inversion process and applying the following

\begin{equation} \label{FinalStokes}
S_s=\mathbf{B^{-1}}\mathbf{T^{-1}}S_m.
\end{equation}

Finally, once $S_s$ is obtained, one or more flux-density calibrators can be used to dermine the K/Jy conversion factor for $I$ and $V$, and a strongly polarized calibrator can be used to determine the K/Jy conversion factor for $Q$ and $U$.

%Ch. 6
\section{Observed results}
%Ch. 6.1
\subsection{Instrumental terms}
During the year 2007 several observations were carried out at the 100-m Effelsberg telescope to test the new sub-reflector. Part of this test time has also been used to test the full Stokes polarimetric calibration. A selected sample of 43 sources (39 extragalactic sources, 2 planetary nebulae, 2 planets) have been observed nearly monthly at 6 cm. Subsequently, some observations at 11 cm, 3.6 cm and 2.8 cm were also carried out.

The Stokes parameters measurement listed in Table \ref{t:pol6} for the planetary nebula NGC\,7027 yielded also the following instrumental terms (averages over time, and standard deviations)
\begin{align*}
&D_L = (0.59 \pm 0.04) \% \\ 
&D_R = (0.16 \pm 0.08) \% \\
&\varphi_L = (-2.4 \pm 3.8)^{\circ} \\
&\varphi_R = (-3.7 \pm 8.4)^{\circ} \\
&\Psi = (-4.1 \pm 2.7)^{\circ}.
\end{align*}

These values are in good agreement with the receiver technical specifications. The measured amplitudes are in good agreement also with EVN measurements. The definition of D-terms phase in the interferometry processing is quite obscure, so it is unclear whether the results are different or not.

\begin{table*}[htbp]
\caption{Observed Stokes parameters for NGC\,7027 at 6 cm. Observations were made in 2007 at Effelsberg. Units are kelvin}
\label{t:pol6}
\centering
\begin{tabular}{lrrrrrrrr}
\hline
Month & I & $\sigma$ & Q & $\sigma$ & U & $\sigma$ & V & $\sigma$ \\
\hline
February & 17.014 & 0.143 & -0.079 & 0.004 & -0.016 & 0.006 & -0.106 & 0.018 \\
April & 17.438 & 0.289 & -0.070 & 0.002 & -0.015 & 0.005 & -0.133 & 0.028 \\
May & 16.864 & 0.079 & -0.073 & 0.005 & -0.020 & 0.008 & -0.108 & 0.028 \\
June & 16.989 & 0.061 & -0.066 & 0.004 & -0.018 & 0.008 & -0.122 & 0.019 \\
August & 17.179 & 0.245 & -0.066 & 0.002 & -0.015 & 0.007 & -0.124 & 0.017 \\
November & 17.068 & 0.343 & -0.076 & 0.011 & -0.014 & 0.015 & -0.087 & 0.022 \\
December & 17.279 & 0.265 & -0.075 & 0.010 & -0.019 & 0.015 & -0.093 & 0.025 \\
\hline
\end{tabular}
\end{table*}

%Ch. 6.2
\subsection{Circular polarization}
We used our D-term calibration to search for CP in 19 AGNs with improved sensitivity over previous observations at Effelsberg.

Table \ref{t:CP} lists a comparison at 6 cm between the raw circular polarization and the values obtained after the calibration, using NGC\,7027 as unpolarized calibrator and 3C\,286 as strongly polarized calibrator. The data were observed in November 2007.

After the D-term calibration, the second planetary nebula, NGC\,6572, appears circularly unpolarized as expected. Significant amounts of CP were detected in the two known CP sources 0743-006 and 1519-273 at levels consistent with published levels (0743-006: $V=(-0.51 \pm 0.08) \%$, \citet{Allers} and $V = (-0.46 \pm 0.05) \%$, \citet{Homan}, both adopting the circular RL frame and for 1519-273: $V = (-0.92 \pm 0.17) \%$ \citep{Allers}). In the Effelsberg measurement of 1519-273, the source appeared unpolarized from the raw values, but the D-terms calibration revealed a level of CP in good agreement with the values reported in the literature. The Effelsberg measurement of the weakly circularly polarized source 3C\,279 also shows a level of CP close to that published ($V =(-0.17 \pm 0.01) \%$ \citep{Allers}). 

\begin{table*}[htbp]
\caption{Observed circular polarization, before and after applying the D-terms calibration}
\label{t:CP}
\centering
\begin{tabular}{llrrrrrrrr}
\hline
Source & & $I_m$ & $\sigma$ & $V_m$ & $\sigma$ & $I_s$ & $\sigma$ & $V_s$ & $\sigma$ \\
 & & [K] & & [\%] & & [K] & & [\%] & \\
\hline
0056$-$00 & & 4.180 & 0.045 & -0.09 & 0.16 & 4.193 & 0.083 & 0.28 & 0.12 \\
0134+32 & 3C\,48 & 17.218 & 0.183 & -0.19 & 0.10 & 17.288 & 0.266 & 0.24 & 0.09 \\
0153+74 & & 3.134 & 0.032 & -0.03 & 0.20 & 3.147 & 0.036 & 0.42 & 0.04 \\
0316+41 & 3C\,84 & 46.456 & 0.478 & -0.44 & 0.11 & 46.651 & 0.655 & 0.00 & 0.06 \\
0212+73 & & 12.913 & 0.120 & -0.36 & 0.11 & 12.967 & 0.146 & 0.08 & 0.04 \\
0607$-$15 & & 11.352 & 0.148 & -0.16 & 0.07 & 11.398 & 0.201 & 0.28 & 0.08 \\
0716+71 & & 2.991 & 0.028 & -0.28 & 0.16 & 3.004 & 0.035 & 0.18 & 0.04 \\
0743$-$00 & & 6.063 & 0.015 & -0.09 & 0.10 & 6.089 & 0.132 & 0.36 & 0.08 \\
0835+58 & & 1.867 & 0.027 & -0.21 & 0.25 & 1.880 & 0.031 & 0.25 & 0.12 \\
0836+71 & & 6.692 & 0.069 & -0.37 & 0.15 & 6.720 & 0.085 & 0.09 & 0.08 \\
0951+69 & 3C\,231 & 10.320 & 0.128 & -0.20 & 0.15 & 10.362 & 0.117 & 0.25 & 0.04 \\
1226+02 & 3C\,273 & 117.944 & 1.083 & -0.43 & 0.09 & 118.458 & 2.049 & 0.04 & 0.09\\
1253$-$05 & 3C\,279 & 36.672 & 0.223 & -0.33 & 0.11 & 36.831 & 0.635 & 0.11 & 0.07 \\
1328+30 & 3C\,286 & 23.331 & 0.305 & -0.28 & 0.11 & 23.428 & 0.307 & 0.14 & 0.11 \\
1409+52 & 3C\,295 & 20.688 & 0.439 & -0.20 & 0.11 & 20.763 & 0.357 & 0.25 & 0.07 \\
1519$-$27 & & 4.700 & 0.079 & 0.20 & 0.21 & 4.719 & 0.114 & 0.62 & 0.15 \\
1749+09 & & 10.873 & 0.097 & -0.37 & 0.12 & 10.919 & 0.238 & 0.07 & 0.10 \\
1800+44 & & 3.830 & 0.079 & -0.49 & 0.12 & 3.844 & 0.044 & -0.04 & 0.05 \\
2134+00 & & 31.723 & 0.672 & -0.45 & 0.07 & 31.836 & 0.551 & -0.01 & 0.07 \\
2145+06 & & 18.925 & 0.211 & -0.39 & 0.07 & 19.003 & 0.284 & 0.05 & 0.06 \\
NGC\,6572& & 3.861 & 0.008 & -0.42 & 0.11 & 3.878 & 0.117 & 0.03 & 0.12 \\
NGC\,7027& &17.068 & 0.343 & -0.51 & 0.13 & 17.130 & 0.218 & -0.06 & 0.05 \\
SATURN  & & 1.962 & 0.028 & -0.85 & 0.19 & 2.165 & 0.042 & -0.30 & 0.09 \\
\hline
\end{tabular}
\end{table*}

Time variability in CP and changes of sign are very common \citep{Allers}, 1519-273 is the most stable CP source known so far. Table \ref{t:CPTime} shows a sample of CP values that we observed during 2007. Our detections also confirm the stability of the CP level of 1519-273.

\begin{table*}[htbp]
\caption{Repeated observations of circular polarization, in percentage, from a sample of sources, during 2007.}
\label{t:CPTime}
\centering
\begin{tabular}{llrrrrrrrrrrrrrr}
\hline
 & & Feb & & Apr & & May & & Jun & & Aug & & Nov & & Dec & \\
\hline
Source & & V & $\sigma$ & V & $\sigma$ & V & $\sigma$ & V & $\sigma$ & V & $\sigma$ & V & $\sigma$ & V & $\sigma$ \\
\hline
0743$-$00 & & 0.53 & 0.04 & 0.42 & 0.05 & 0.24 & 0.08 & 0.18 & 0.06 & 0.52 & 0.04 & 0.36 & 0.08 & 0.38 & 0.06 \\
1253$-$05 & 3C\,279   & 0.11 & 0.05 & -0.12 & 0.08 & -0.40 & 0.09 & -0.25 & 0.06 & -0.35 & 0.10 & 0.11 & 0.07 & 0.29 & 0.12 \\
1519$-$27 & & - & - & 0.65 & 0.14 & 0.53 & 0.08 & 0.61 & 0.11 & - & - & 0.62 & 0.15 & - & - \\ 
NGC\,6572 & & - & - & -0.33 & 0.19 & -0.06 & 0.04 & - & - & - & - & 0.03 & 0.12 & 0.19 & 0.13  \\
\hline
\end{tabular}
\end{table*}

%Ch. 6.3
\subsection{Linear polarization}
The proposed calibration scheme maintains the same reliability in the calibration of linear polarization as that reported by \citet{Turlo} based on a 3$\times$3 M\"uller matrix. In addition our method supplies Stokes $V$, the D-terms and the complete error treatment.

The traditional polarization calibration is applied assuming a M\"uller matrix unaffected by errors, whereas the scheme that we propose includes the error  propagation also during the matrix inversion and its application, hence the resulting error estimates are intrinsically higher than those obtained with the standard procedure. Tables \ref{t:LP} and \ref{t:PA} show the comparison between the standard and the new calibration schemes, when measuring linear polarization.

The linear polarization values obtained are consistent with those obtained with the standard method.

\begin{table*}[htbp]
\caption{Observed linear polarization, in percentage, from a sample of sources measured at 6 cm at Effelsberg during 2007. Superscripts $^1$ indicates the results obtained with the new calibration procedure, $^2$ those obtained with the traditional calibration procedure.}
\label{t:LP}
\centering
\begin{tabular}{llrrrrrrrrrrrrrr}
\hline
 & & Feb & & Apr & & May & & Jun & & Aug & & Nov & & Dec & \\
\hline
Source & & P & $\sigma$ & P & $\sigma$ & P & $\sigma$ & P & $\sigma$ & P & $\sigma$ & P & $\sigma$ & P & $\sigma$ \\
\hline
0743$-$00 $^1$ & & 0.81 & 0.19 & 0.87 & 0.12 & 0.97 & 0.54 & 1.28 & 0.27 & 1.12 & 0.17 & 1.11 & 0.60 & 0.84 & 0.53 \\
0743$-$00 $^2$ & & 0.82 & 0.13 & 0.85 & 0.09 & 0.97 & 0.20 & 1.24 & 0.06 & 1.06 & 0.08 & 0.98 & 0.09 & 0.90 & 0.13 \\
1253$-$05 $^1$ & 3C\,279  & 3.25 & 0.28 & 1.24 & 0.21 & 0.81 & 0.45 & 0.96 & 0.40 & 1.14 & 0.36 & 1.42 & 0.86 & 1.11 & 1.63 \\
1253$-$05 $^2$ & 3C\,279  & 3.15 & 0.08 & 1.17 & 0.03 & 0.82 & 0.03 & 0.91 & 0.02 & 1.11 & 0.04 & 1.16 & 0.04 & 1.06 & 0.00 \\
1519$-$27 $^1$ & & - & - & 3.62 & 0.71 & 2.96 & 0.25 & 5.79 & 0.58 & - & - & 4.70 & 0.62 & - & - \\ 
1519$-$27 $^2$ & & - & - & 3.50 & 0.01 & 3.20 & 0.16 & 5.83 & 0.01 & - & - & 4.55 & 0.27 & - & - \\ 
NGC\,6572 $^1$ & & - & - & 0.09 & 0.79 & - & - & - & - & - & - & 0.21 & 1.27 & 0.31 & 0.77 \\
NGC\,6572 $^2$ & & - & - & 0.16 & 0.01 & - & - & - & - & - & - & 0.24 & 0.27 & 0.48 & 0.59 \\
\hline
\end{tabular}
\end{table*}

\begin{table*}[htbp]
\caption{Observed polarization angle, in degrees, from a sample of sources measured at 6 cm at Effelsberg during 2007, superscripts $^1$ indicate the results obtained with the new calibration procedure, $^2$ those obtained with the traditional calibration procedure}
\label{t:PA}
\centering
\begin{tabular}{llrrrrrrrrrrrrrr}
\hline
 & & Feb & & Apr & & May & & Jun & & Aug & & Nov & & Dec & \\
\hline
Source & &  PA & $\sigma$ & PA & $\sigma$ & PA & $\sigma$ & PA & $\sigma$ & PA & $\sigma$ & PA & $\sigma$ & PA & $\sigma$ \\
\hline
0743$-$00 $^1$ & & 29.5 & 0.4 & 31.4 & 0.2 & 32.2 & 0.3 & 28.1 & 0.4 & 30.6 & 0.2 & 31.0 & 1.0 & 26.0 & 0.6 \\
0743$-$00 $^2$ & & 28.8 & 1.8 & 30.6 & 2.1 & 31.8 & 1.4 & 29.6 & 1.4 & 29.8 & 1.2 & 34.4 & 2.3 & 24.6 & 7.5 \\
1253$-$05 $^1$ & 3C\,279  & -28.9 & 0.1 & -28.7 & 0.2 & -36.5 & 0.4 & -17.7 & 0.4 & -1.3 & 0.4 & 12.0 & 0.5 & 9.6 & 0.8 \\
1253$-$05 $^2$ & 3C\,279  & -29.9 & 0.6 & -28.8 & 1.3 & -31.6 & 1.7 & -17.9 & 0.6 & -1.9 & 1.0 & 13.5 & 0.5 & 8.6 & 0.1 \\
1519$-$27 $^1$ & & - & - & 3.3 & 0.1 & -34.3 & 0.1 & -36.4 & 0.2 & - & - & 130.1 & 0.3 & - & - \\ 
1519$-$27 $^2$ & & - & - & 2.2 & 0.3 & -28.1 & 6.6 & -36.6 & 0.1 & - & - & 129.4 & 1.8 & - & - \\ 
\hline
\end{tabular}
\end{table*}

%Ch. 7
\section{Conclusions}
A calibration procedure based on a 4$\times$4 M\"uller matrix has been proposed, that allows Stokes $V$ measurements with single-dish telescopes that supply circularly polarized outputs. This method also provides Stokes $I, Q$ and $U$, and so allows a full polarimetric study of sources under examination.

The mathematical interpretation of the M\"uller matrix elements offers the possibility of deriving the instrumental polarimetric characteristics of the telescope (D-terms, in amplitude and phase) directly from the observation, allowing a direct comparison with the technical specifications set by the engineers.

The proposed procedure is valid for all single-dish telescopes equipped with a receiving architecture composed of a standard receiver, two total power detectors for the two circularly polarized channels and a multiplying polarimeter (one of the most common solution available nowadays), and can be applied at any observed frequency. 

%App.A
\appendix	
\section{The backend and the derivation of the measured Stokes parameters}
\label{App1}
In the following, the average symbol $<>$ is omitted, subscript $s$ stands for source, and subscript $c$ for calibration noise diode.

\subsection{Total power detectors}
To measure total power in the R and L polarizations, the measurements $\widetilde V_R\widetilde V_R^*=V_R^2$ and $\widetilde V_L\widetilde V_L^*=V_L^2$ are made. According to Eq. \eqref{conway}, introducing the square law detector gain (the scalar quantity $g$) and recalling that
\begin{align*}
&\left(\widetilde A\widetilde B\right)^*=\widetilde A^*\widetilde B^*\\
&\left(\widetilde A+\widetilde B\right)^*=\widetilde A^*+\widetilde B^*
\end{align*}

\noindent one obtains
\begin{subequations} \label{sldampl}
\begin{align}
\widetilde V_R\widetilde V_R^* & =\frac{g_2G_R^2 }{1 + D_R^2}\left[E_R^2+D_R^2 E_L^2+ \widetilde E_R \widetilde E_L^*\widetilde D_R^*+ \widetilde E_L\widetilde E_R^*\widetilde D_R \right]\\ 
\widetilde V_L\widetilde V_L^* & =\frac{g_1G_L^2 }{1 + D_L^2}\left[E_L^2+D_L^2 E_R^2 +\widetilde E_L \widetilde E_R^*\widetilde D_L^*+ \widetilde E_R\widetilde E_L^*\widetilde D_L \right]
\end{align}
\end{subequations}

\noindent Eq. \eqref{sldampl} can be expressed in terms of the Stokes parameters or in terms of source linearly polarized flux $I_{lps}$ and phase differences. When the Tcal calibration is applied, a common post-processing procedure allows one to transform the voltage at the output of the total power devices (TP) into kelvin, as follows

\begin{subequations} \label{Tcal}
\begin{align}
TP_R & = \frac{V_{Rs}^2}{V_{Rc}^2}\frac{I_c}{2}\\
TP_L & = \frac{V_{Ls}^2}{V_{Lc}^2}\frac{I_c}{2}\
\end{align}
\end{subequations}

The Tcal is usually fully linearly polarized and characterized by $V_{c}=0 K$. This simplifies the involved calculations.
By comparing the measured signal with that of the noise diode, the receiver gain fluctuations are removed. In addition, the Tcal calibration procedure calibrates the source strength measurements in terms of antenna temperatures, in kelvin. Discarding the terms of order $\geq$ 3 one obtains

\begin{subequations} \label{iv}
\begin{align}
I_m = & \frac{1}{DN_{IV}}\left\lbrace  I_s [1 + D_R^2 + D_L^2 + D_L \cos (\varphi _L + \Delta _c) \right. \nonumber\\
& \left. + D_R \cos (\varphi _R - \Delta _c )]  \right. \nonumber\\ 
& \left. + Q_S [D_R \cos \varphi _R + D_L \cos \varphi _L + 2D_R D_L \cos \varphi _R \cos (\varphi _L + \Delta _c) \right. \nonumber\\
& \left. + 2D_R D_L \cos \varphi _L \cos (\varphi _R - \Delta _c)]  \right. \nonumber\\ 
& \left. + U_S [D_R \sin \varphi _R + 2D_R D_L \sin \varphi _R \cos (\varphi _L  + \Delta _c) - D_L \sin \varphi _L \right. \nonumber\\
& \left. - 2D_R D_L \sin \varphi _L \cos (\varphi _R - \Delta _c)] \right. \nonumber\\ 
& \left. + V_s [D_R^2 - D_L^2 - D_L \cos (\varphi _L + \Delta _c) + D_R \cos (\varphi _R - \Delta _c)] \right\rbrace \\ 
V_m = & \frac{1}{DN_{IV}}\left\lbrace I_s [D_R \cos (\varphi _R - \Delta _c) - D_L \cos (\varphi _L + \Delta _c)] \right. \nonumber\\ 
& \left. + Q_S [D_L\cos \varphi _L + 2D_R D_L \cos \varphi _L \cos (\varphi _R - \Delta _c) - D_R \cos \varphi _R \right. \nonumber\\
& \left.- 2D_R D_L \cos \varphi _R \cos (\varphi _L + \Delta _c)] \right. \nonumber\\ 
& \left. + U_S [-D_L \sin \varphi _L - 2D_R D_L \sin \varphi _L \cos (\varphi _R - \Delta _c) - D_R \sin \varphi _R \right. \nonumber\\
& \left. - 2D_R D_L \sin \varphi _R \cos (\varphi _L + \Delta _c)] \right. \nonumber\\ 
& \left. + V_s [1 + D_R \cos (\varphi _R - \Delta _c) + D_L \cos (\varphi _L + \Delta _c)] \right\rbrace 
\end{align}
\text{where}
\begin{align*}
DN_{IV} = & 1 + D_R^2 + D_L^2 + 2D_R \cos (\varphi _R  - \Delta _c ) + 2D_L \cos (\varphi _L + \Delta _c)\\
& + 4D_R D_L \cos (\varphi _R - \Delta _c )\cos (\varphi _L + \Delta _c).
\end{align*}
\end{subequations}

These two equations show how the measured $I_m$ and $V_m$ are contaminated by $I_s, Q_s, U_s$ and $V_s$ depending on the values of the D-terms.

\subsection{Multiplying polarimeter}

To measure the linear polarization components, $Q_m$ and $U_m$, the measurements $\widetilde V_R \widetilde V_L^*+\widetilde V_L\widetilde V_R^*$ and $i\left(\widetilde V_R\widetilde V_L^*-\widetilde V_L\widetilde V_R^*\right) $ are made using a multiplying polarimeter. The outputs of the multipliers in the polarimeter are related to the incoming electric fields by

\begin{subequations} \label{crosscor}
\begin{align}
\widetilde V_R\widetilde V_L^* & = \frac{\widetilde G_R\widetilde G_L^*}{DN_{MP}}\left[\widetilde E_R\widetilde E_L^*+ \widetilde D_R 
  E_L^2 + \widetilde D_L^* E_R^2 + \widetilde D_R\widetilde D_L^* \widetilde E_L \widetilde E_R^*\right] \\ 
\widetilde V_L\widetilde V_R^* & = \frac{\widetilde G_L\widetilde G_R^*}{DN_{MP}}\left[\widetilde E_L\widetilde E_R^*+ \widetilde D_R^*
  E_L^2 + \widetilde D_L E_R^2 + \widetilde D_R^* \widetilde D_L \widetilde E_L^* \widetilde E_R\right]
\end{align}
\text{where}
\begin{align*}
DN_{MP} = & \sqrt{1+D_R^2+D_L^2+D_R^2 D_L^2}.
\end{align*}
\end{subequations}

\noindent The Eqs. \eqref{crosscor} can be expressed in terms of Stokes parameters or in terms of linearly polarized flux density.
By assuming that $g_Q=g_U$ and $\gamma_Q=\gamma_U$ (standard procedure at Effelsberg), it is possible to apply to both the Q and U channels the same calibration factor, according to the following
\begin{equation} \label{cal}
I_{lpc}=\sqrt{{Q_c}^{2}+{U_c}^{2}}
\end{equation}

\noindent where $Q_c$ and $U_c$ are the Tcal signals coming through the $Q$ and $U$ channels and measured at the end of the backend. 
\\

The Tcal is applied analogously to Eqs. \eqref{Tcal}. The ratio between the $I_{lpc}$ value and the corresponding measured voltage gives the conversion factor $K/V$ to be applied to the on-source $Q$ and $U$ measurements

\begin{subequations} \label{QUTcal}
\begin{align}
Q_m & = \frac{Q_s}{\sqrt{{Q_c}^{2}+{U_c}^{2}}}\cdot I_{lpc}\\
U_m & = \frac{U_s}{\sqrt{{Q_c}^{2}+{U_c}^{2}}}\cdot I_{lpc}\
\end{align}
\end{subequations}

Discarding the terms of order $\geq$ 3 the following is obtained

\begin{subequations} \label{qu}
\begin{align}
Q_m  = & \frac{1}{DN_{QU}} \left\lbrace I_s \left[D_L \cos \left(\Psi  + \varphi _L \right) + D_R \cos \left(\Psi  - \varphi _R \right) \right] \right. \nonumber\\ 
& \left. + Q_s \left[\cos \Psi + D_R D_L \cos \left(\Psi + \Phi \right) \right] \right. \nonumber\\ 
& \left. + U_s \left[\sin \Psi - D_R D_L \sin \left(\Psi + \Phi \right) \right] \right. \nonumber\\ 
& \left. + V_s \left[D_R \cos \left(\Psi - \varphi _R \right) - D_L \cos \left(\Psi + \varphi _L \right) \right] \right\rbrace\\ 
\nonumber\\
U_m = & \frac{1}{DN_{QU}} \left\lbrace I_s \left[-D_L \sin \left(\Psi + \varphi _L \right) - D_R \sin \left(\Psi - \varphi _R \right) \right] \right. \nonumber\\
& \left. + Q_s \left(-\sin \Psi - D_R D_L \sin \left(\Psi + \Phi \right) \right] \right. \nonumber\\
& \left. + U_s \left(\cos \Psi - D_R D_L \cos \left(\Psi + \Phi \right) \right] \right. \nonumber\\ 
& \left. + V_s \left(-D_R \sin \left(\Psi - \varphi _R \right) + D_L \sin \left(\Psi + \varphi _L \right) \right] \right\rbrace
\end{align}
\text{where}
\begin{align*}
DN_{QU} = & \left[1 + D_L^2 + D_R^2 + 2 D_L D_R \cos\left(\Phi+\Delta \right) \cos \Delta \right.\nonumber\\
& \left. + 2 D_L \cos \left( \varphi_L + \Delta \right) + 2 D_R \cos \left( \Delta - \varphi_R \right)\right]^{\frac{1}{2}}.
\end{align*}
\end{subequations}

The channels Q and U could also be calibrated separately. In this case in Eqs. \eqref{qu} two different denominators, one for each channel, would be present.

With Eqs. \eqref{iv} and \eqref{qu} we have all 16 elements of the M\"uller matrix in terms of D-terms, required to relate measured and true Stokes parameters.

%App. B
\section{Definitions of the matrix elements}
\label{App2}
From Eqs. \eqref{iv}, \eqref{qu} and recalling the definition \eqref{mueller1}, the coefficients of the instrumental M\"uller matrix $\mathbf{T}$ can be summarized as follows:\\

\begin{subequations} \label{dlist}
\noindent \text{$\bullet$ Propagation of $I_{s}$ into $I_{m}$}
\begin{equation}
t_{11} = \frac{1}{DN_{IV}} \left[1 + D_R^2 + D_L^2 + D_L \cos (\varphi _L  + \Delta _c ) + D_R \cos (\varphi _R  - \Delta _c )\right]
\end{equation}

\noindent \text{$\bullet$ Propagation of $Q_{s}$ into $I_{m}$}
\begin{align}
t_{12} = &\frac{1}{DN_{IV}}\left[D_R\cos \varphi _R + D_L\cos \varphi _L + 2D_R D_L \cos \varphi _R \cos (\varphi _L + \Delta _c) \right.\nonumber\\
& \left. + 2D_R D_L \cos \varphi _L \cos (\varphi _R - \Delta _c )\right]\
\end{align}

\noindent \text{$\bullet$ Propagation of $U_{s}$ into $I_{m}$}
\begin{align}
t_{13}= &\frac{1}{DN_{IV}} \left[D_R \sin \varphi _R + 2D_R D_L \sin \varphi _R \cos (\varphi _L + \Delta _c ) - D_L \sin \varphi _L \right. \nonumber\\
& \left. - 2D_R D_L \sin \varphi _L \cos (\varphi _R  - \Delta _c )\right]\
\end{align}

\noindent \text{$\bullet$ Propagation of $V_{s}$ into $I_{m}$}
\begin{equation}
t_{14}=\frac{1}{DN_{IV}}\left[D_R^2 - D_L^2 - D_L \cos (\varphi _L + \Delta _c) + D_R \cos (\varphi _R - \Delta _c )\right]
\end{equation}

\noindent \text{$\bullet$ Propagation of $I_{s}$ into $Q_{m}$}
\begin{equation}
t_{21}=\frac{1}{DN_{QU}}\left[D_L \cos \left(\Psi + \varphi _L \right) + D_R \cos \left(\Psi - \varphi _R \right) \right] 
\end{equation}

\noindent \text{$\bullet$ Propagation of $Q_{s}$ into $Q_{m}$}
\begin{equation}
t_{22}=\frac{1}{DN_{QU}}\left[\cos \Psi + D_R D_L \cos \left(\Psi + \Phi \right) \right]
\end{equation}

\noindent \text{$\bullet$ Propagation of $U_{s}$ into $Q_{m}$}
\begin{equation}
t_{23}=\frac{1}{DN_{QU}}\left[\sin \Psi - D_R D_L \sin \left(\Psi + \Phi \right) \right]
\end{equation}

\noindent \text{$\bullet$ Propagation of $V_{s}$ into $Q_{m}$}
\begin{equation}
t_{24}=\frac{1}{DN_{QU}}\left[D_R \cos \left(\Psi - \varphi _R \right) - D_L \cos \left(\Psi + \varphi _L \right) \right]
\end{equation}

\noindent \text{$\bullet$ Propagation of $I_{s}$ into $U_{m}$}
\begin{equation}
t_{31}=\frac{1}{DN_{QU}}\left[-D_L \sin \left(\Psi + \varphi _L \right) - D_R \sin \left(\Psi - \varphi _R \right) \right]
\end{equation}

\noindent \text{$\bullet$ Propagation of $Q_{s}$ into $U_{m}$}
\begin{equation}
t_{32}=\frac{1}{DN_{QU}}\left[-\sin \Psi - D_R D_L \sin \left(\Psi + \Phi \right) \right]
\end{equation}

\noindent \text{$\bullet$ Propagation of $U_{s}$ into $U_{m}$}
\begin{equation}
t_{33}=\frac{1}{DN_{QU}}\left[\cos \Psi  - D_R D_L \cos \left(\Psi + \Phi \right) \right]
\end{equation}

\noindent \text{$\bullet$ Propagation of $V_{s}$ into $U_{m}$}
\begin{equation}
t_{34}=\frac{1}{DN_{QU}}\left[-D_R \sin \left(\Psi - \varphi _R \right) + D_L \sin \left(\Psi + \varphi _L \right) \right]
\end{equation}

\noindent \text{$\bullet$ Propagation of $I_{s}$ into $V_{m}$}
\begin{equation}
t_{41}=\frac{1}{DN_{IV}}\left[D_R \cos (\varphi _R - \Delta _c ) - D_L \cos (\varphi _L + \Delta _c )\right] 
\end{equation}

\noindent \text{$\bullet$ Propagation of $Q_{s}$ into $V_{m}$}
\begin{align}
t_{42}=&\frac{1}{DN_{IV}}\left[D_L \cos \varphi _L + 2D_R D_L \cos \varphi _L \cos (\varphi _R - \Delta _c ) - D_R \cos \varphi _R \right. \nonumber\\
&\left. - 2D_R D_L \cos \varphi _R \cos (\varphi _L + \Delta _c )\right]\
\end{align}

\noindent \text{$\bullet$ Propagation of $U_{s}$ into $V_{m}$}
\begin{align}
t_{43}=&\frac{1}{DN_{IV}}\left[-D_L \sin \varphi _L - 2D_R D_L \sin \varphi _L \cos (\varphi _R - \Delta _c ) - D_R \sin \varphi _R \right. \nonumber\\
&\left. - 2D_R D_L \sin \varphi _R \cos (\varphi _L + \Delta _c )\right]\
\end{align}

\noindent \text{$\bullet$ Propagation of $V_{s}$ into $V_{m}$}
\begin{equation}
t_{44}=\frac{1}{DN_{IV}}\left[1 + D_R \cos (\varphi _R - \Delta _c) + D_L \cos (\varphi _L + \Delta _c )\right] 
\end{equation}
\end{subequations}

\begin{acknowledgements}
This research was supported by the EU Framework 6 Marie Curie Early Stage Training programme under contract number MEST-CT-2005-19669 ``ESTRELA".
This work is based on observations with the 100-m telescope of the MPIfR (Max-Planck-Institut f\"ur Radioastronomie) at Effelsberg. Effelsberg D-terms, coming from EVN measurements at 6 cm, were kindly given us by Simone Bernhart. Thanks to Alan Roy for the careful review, the detailed suggestions and the language corrections. Thanks to Johan P. Hamaker, who refereed the paper, for the interesting comments. Finally, it must be mentioned that this paper has been written under the constant and $quiet$ supervision of someone $extremely$ $young$.
\end{acknowledgements}

\bibliographystyle{aa}
\bibliography{Ref3.bib}

\end{document}